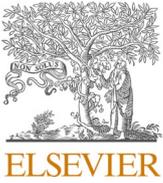
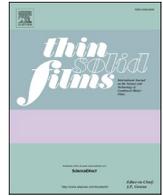

# Evolution of local atomic structure accompanying devitrification of amorphous Ni-Zr alloy thin films

Debarati Bhattacharya [a,c,*], Nidhi Tiwari [b], P.S.R. Krishna [a,c], Dibyendu Bhattacharyya [b,c]

[a] *Solid State Physics Division, Bhabha Atomic Research Centre, Mumbai 400085, India*
[b] *Atomic and Molecular Physics Division, Bhabha Atomic Research Centre, Mumbai 400085, India*
[c] *Homi Bhabha National Institute, Anushaktinagar, Mumbai 400094, India*

ABSTRACT

Thin film metallic glasses undergoing devitrification can form partially crystallized or fully crystallized materials with novel structural and magnetic properties. The development of desired and tunable properties of such systems drives the need to understand the mechanism of their thermal evolution at the atomic level. Co-sputtered amorphous Ni-Zr alloy thin films which were thermally annealed in steps of 200°C from room temperature up to 800°C, were observed to undergo an amorphous-to-crystalline transformation. Evolutions in local atomic structure, including oxide formation and depletion during this devitrification process, were determined using a combination of X-ray reflectivity (XRR), grazing incidence X-ray diffraction (GIXRD) and Extended X-ray Absorption Fine Structure (EXAFS) techniques. The most probable phase of the alloy was determined as $Ni_7Zr_2$; undergoing a polymorphous transformation. The slight oxide content in the films was also noted to decrease rapidly as annealing proceeded; leaving the crystallization pathway unhindered. modeling and analyses of XRR and GIXRD data showed that film thickness decreased with annealing while long range order increased. Pair distribution function peak widths were observed to decrease with annealing, indicating a control over structural disorder in the system as it transitioned from the amorphous to crystalline state. Partial atomic distribution in the environment of each constituent was examined at every stage of annealing through EXAFS measurements, which gave proper insight into atomic scale changes to the order present. Detailed analyses through these techniques gave coherent results, thus providing a true picture of the thermal evolution process of devitrification in this technologically useful material.

## 1. Introduction

Amorphous metal alloys in thin film form or thin film metallic glasses (TFMGs) are known to possess exceptional properties leading to their potential use in various applications [1]. TFMGs exhibit chemical, mechanical and thermal properties superior to their bulk counterparts [2], allowing them to be potentially useful in many practical applications: as in structural engineering [3], as functional materials [4,5] and also for enhancing the performance of medical tools [6,7]. A fundamental feature associated with most of these TFMG applications is high mechanical strength [8], which is closely related to the chemical and structural order as well as glass forming ability of these amorphous materials [9]. Metallic glasses have an absence of long range order but they are suitably characterized by their inherent short range order [10]; which arises from the strong interatomic bonding between their constituent elements [11]. Several recent reports through both experimental and simulated studies, elucidate the important aspect of local atomic structure analysis to engender a deeper understanding of the structural properties of metallic glasses [12–18]. Investigation of local atomic structure in metallic glasses is further crucial to the understanding of their structure–property relationships; in particular, their mechanical properties as evinced by relevant literature [9,19–21].

The viability of functional materials is often put to test during high temperature applications and possible exposure to oxidizing environments. It is due to this aspect that studies regarding thermal stability of TFMGs merit further investigation. Devitrification or crystallization of the glassy state occurs in TFMGs when subjected to thermal treatments; which can create improved properties [22–24] just as is the case for nanostructuring reported in heated bulk amorphous alloys [25,26]. This leads to the formation of either pure or mixed phases in the annealed thin films. Thus TFMGs can also act as precursors for such partially crystallized or fully crystallized forms which display novel structural [27], mechanical [28,29] and magnetic properties [30]. The development of desired and tunable properties in these materials drives the need to understand the atomistic mechanism of thermal evolution of such systems. Information regarding local atomic structure through short range order studies, which defines the variation of atomic environment in the neighborhood of surrounding atoms, also contributes to a better understanding of structural relaxation in metallic glasses [31–34]. In some cases it has been observed that preferred devitrification pathways can unexpectedly change as a result of oxidation in both bulk metallic glass [35,36] and TFMG [37] systems, while others report the presence of oxygen producing favourable metallic glass properties [38,39]. Investigation of local atomic structure in these materials can also determine the oxygen sites in the atomic environment under study.

The binary Ni-Zr system is known to have strong interaction between the two transition metal species, the atomic size difference notwithstanding, and forms Ni-Zr glassy alloys over a wide composition range. It is obvious that these phase formations are strongly dictated by thermodynamics [40], aided by the fact that highly negative enthalpies accompany the formation of each phase. Structural relaxation and crystallization dynamics in these materials have been studied through short range studies to establish a link between the glassy and crystalline states [41]. Ni-rich bulk glassy alloys have been studied for their specific thermal properties and behavior during crystallization; as in the formation of intermediate phases during the process [42]. Further, Ni-rich TFMGs of the Ni-Zr system have been studied for their thermal properties with regard to their important contribution towards hydrogen separation applications [43,44]. Ni-Zr TFMGs have also been explored for their mechanical properties [21,45,46]. Out of the Ni-rich phases of the Ni-Zr system, the congruently melting $Ni_7Zr_2$ phase has the poorest glass forming ability [47]. Thus this phase provides ample scope to study its mechanism of crystallization in a TFMG form through atomistic structural changes. In addition, not many studies have been devoted to the structure, formation and transition of this particular phase [30,47–51,47], thereby extending the opportunities of related research. The current work focuses on the thermal evolution of this phase in TFMG form vis-à-vis the changes of its atomic scale structure. Investigations of the structure of Ni-rich Ni-Zr bulk glassy alloys have revolved around the vital role played by short range order in these amorphous systems [9,16,17,52,53]. While a lot of reported work has been devoted to this subject, no consensus exists as yet, thus offering more opportunities for further research and analysis. Independent methods include evaluation of total structure factor from X-ray [54] and neutron [55] scattering measurements and estimation of short range order parameter from Extended X-ray Absorption Fine Structure (EXAFS) data analyses [56]. Another approach was a combination of both these techniques, which complemented each other to give coherent results for bulk Ni-rich Ni-Zr alloys [53,57]. We have applied this combinatorial scheme to uniquely map the thermal evolution of a Ni-rich Ni-Zr TFMG on an atomic scale by studying the development of short range order as a function of annealing temperature.

This paper presents results obtained on examination of the local atomic changes occurring during devitrification of glassy Ni-Zr alloy thin films using X-ray reflectivity (XRR), grazing incidence X-Ray diffraction (GIXRD) and EXAFS techniques as investigative probes. The partial atomic distribution in the environment of each constituent was examined at every stage of annealing which gave suitable insight into atomic scale changes.

## 2. Experimental

Amorphous Ni-Zr alloy films were deposited by co-sputtering from pure Ni and Zr targets on chemically cleaned Si substrates in a direct current (D.C.) magnetron sputtering system. Details of the experimental set up can be found elsewhere [58]. The 3-inch diameter sputter targets used had purity of ~99.95% and were pre-sputtered before use. Since they were arranged in a sputter-up confocal geometry, the substrates were rotated above them to enable uniform deposition from the targets at ~75 mm below. The substrates were inserted into the vacuum chamber through a load lock system to minimize contamination. The base vacuum was of the order of $1.33 \times 10^{-5}$ Pa and the system was maintained at 4.4 Pa during sputter deposition with a dynamic flow of Ar gas. D.C. powers of 47 W and 106 W were delivered to the Ni and Zr targets respectively, during deposition of the alloy films by co-sputtering. The as-deposited films at room temperature (RT) were isochronally annealed in vacuum at intervals of 200 °C up to 800 °C for 1 h each. Annealing of the films was carried out in a vacuum chamber equipped with a substrate heater and thermocouple arrangement. A digital temperature controller was used to monitor the temperature of the heater and the overshoot was within ± 5°C. During annealing the vacuum level in the chamber was $\sim 3 \times 10^{-4}$ Pa. The heating rate employed was 20 °C/min and following the required heat treatment, the films were allowed to cool naturally to ambient temperature while in vacuum. After each anneal, the thin films were characterized through XRR, GIXRD and EXAFS techniques.

The XRR and GIXRD measurements were both made in parallel beam geometry using a rotating anode based Rigaku diffractometer with Cu $K_\alpha$ radiation. In this instrument the sample position is fixed and scanning is achieved by synchronized movement of the source and detector units. Care was taken so that the same critical alignment was used for both measurement modes to enable standard normalization procedures for each film. Thicknesses of the films were estimated through model fitting of X-ray reflectivity data. Simulation of the profiles was carried out using Parratt's formalism [59]. EXAFS measurements on the Ni-Zr alloy films were carried out at the Energy Scanning EXAFS beamline (BL-9) of the Indus-2 Synchrotron Source (2.5 GeV, 100 mA), Raja Ramanna center for Advanced Technology, Indore, India [60]. The data were recorded in fluorescence mode using a VORTEX-make single channel Si drift detector for collection of the total fluorescence signal and an ionization chamber for measuring the incident flux.

## 3. Results

### 3.1. X-Ray reflectivity of Ni-Zr alloy thin films

Specularly reflected X-ray intensities at grazing incidence angle θ on the sample surface were recorded as a function of momentum transfer $q = (4\pi/\lambda)\sin\theta$ in $\text{Å}^{-1}$ (where λ is the wavelength of incident beam) normal to the reflecting surface. The XRR data were fitted with a model of depth dependent electron scattering length density (ESLD) averaged over lateral dimensions of the entire sample. ESLD is a function of number density of the atomic components and has units of $\text{Å}^{-2}$. Fig. 1(a) to (d) show XRR data and fits for the Ni-Zr alloy thin films annealed at various temperatures. The corresponding ESLD depth profiles derived for the annealed films are shown as insets in these figures. After suitable modeling of the XRR data, the average composition of all the films was roughly estimated as Ni: 76.2 at% and Zr: 23.7 at%. Atomic fractions of the binary alloy thin films were calculated using the methodology reported earlier [61]. The Ni-Zr alloy films had thus undergone a polymorphous transformation with annealing.

The nearest Ni-Zr phase identifiable to the observed Ni:Zr ratio of 3.2 corresponded to $Ni_7Zr_2$. It was also observed that this pure crystalline phase existed only in the final annealed film at 800 °C. The XRR data of the alloy thin films annealed at 200 °C, 400 °C and 600 °C, could be fitted only on consideration of a very thin oxide layer ($ZrO_2$ and/or NiO) in addition to the main alloy layer composition. It is known that the Ni-Zr system is quite prone to oxidation [62,63]. The XRR fitting parameters used and resultant composition of each annealed film are listed in Table 1. It was observed that the contribution of this oxide layer diminished with increasing annealing temperature, and was absent in the film annealed at 800 °C.

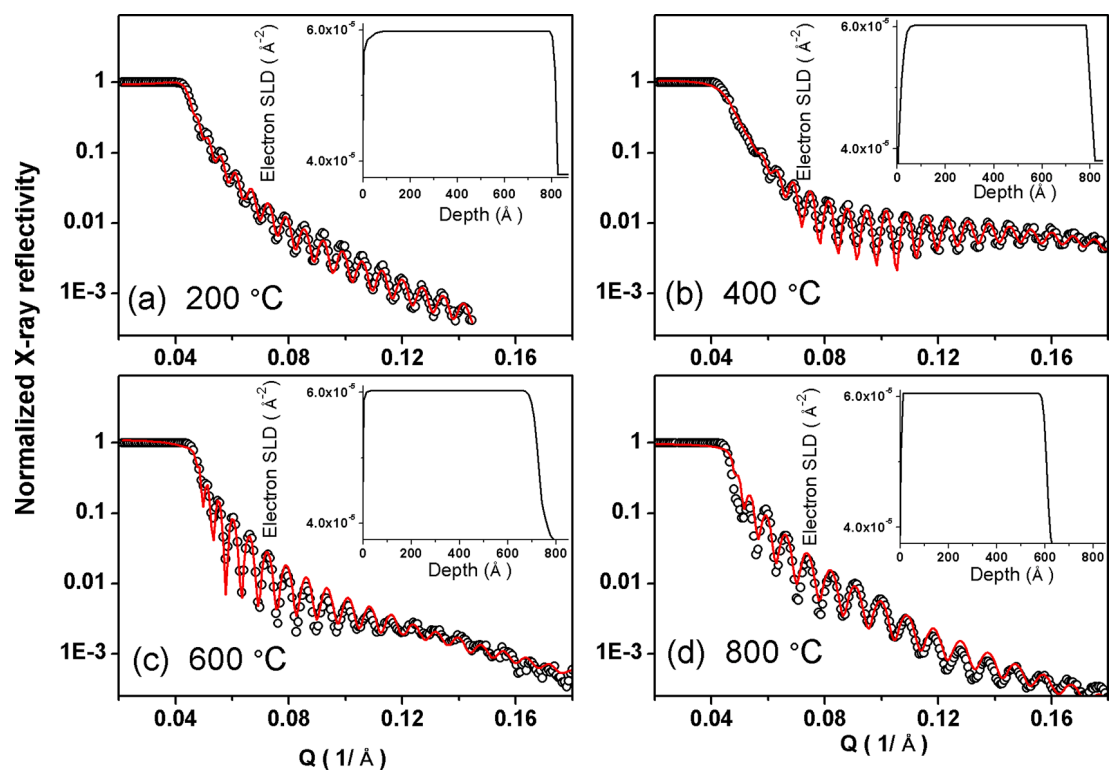

**Fig. 1.** X-ray reflectivity analyses of Ni-Zr thin alloy films thermally annealed at (a) 200 °C (b) 400 °C (c) 600 °C and (d) 800 °C. The data points are represented by open circles and the solid lines are the best fits to the data. The insets are electron scattering length density profiles of the corresponding annealed thin films, showing changes in electron density and thickness.

**Table 1**
XRR fitting parameters for the annealed Ni-Zr alloy films. The films annealed at 200 °C, 400 °C and 600 °C were seen to consist of an alloy layer and a very thin oxide layer, whereas the oxide layer was absent in the alloy film at 800 °C.

| Temperature (C) | Ni-Zr alloy layer | | | Oxide layer | |
|---|---|---|---|---|---|
| | Thickness (Å) | Roughness (Å) | Composition (at%) | Thickness (Å) | Roughness (Å) |
| 200 | 800 | 31.5 | Ni73.6Zr26.4 | 20 | 2.2 |
| 400 | 774 | 15.5 | Ni76.4Zr23.6 | 19 | 2.0 |
| 600 | 730 | 16.0 | Ni76.9Zr23 | 13 | 1.7 |
| 800 | 607 | 3.6 | Ni78Zr22 | – | – |

It can be seen that annealing had affected the thickness of the films. While the reduction in thickness was slight on annealing from 200 °C to 400 °C; the film thickness decreased sharply with further annealing. All the films underwent strong densification with annealing as seen in the difference in vertical heights of the ESLD profiles. The occurrence of increased density with annealing of Ni-Zr glassy films along with volumetric contraction has also been seen earlier [30].

*3.2. Whole pattern analysis of grazing incidence X-Ray diffraction data*

Grazing incidence XRD (GIXRD) data for the Ni-Zr alloy thin films before and after annealing were performed at an incident angle of 0.5°. The diffraction patterns indicated that the films annealed up to 600 °C remained amorphous, while annealing at 800 °C caused transformation into crystalline state. The sharp diffraction peaks in this case could be identified as corresponding to the $Ni_7Zr_2$ stable phase. This is shown in Fig. 2 wherein the peak positions matching the crystallographic phase have been marked. Comparison of the GIXRD patterns of RT deposited and films annealed from 200 °C to 600 °C are shown as an inset to this figure.

In order to obtain more information from the amorphous nature of these films, a whole pattern analysis approach was adopted, not unlike that used to determine the structure of glassy Ni-Zr alloys [54]. This analytical technique utilizes quantitative information from the total pattern so that ideal fitting can proceed with a chosen structure. In this method, analytical line-profile functions and a suitable background model is fitted to the data. A specific advantage of using this method allows the determination of mixed phases and thereby any deviation from the pure phase present. Amorphous structures are characterized by lack of long-range order. However, their local atomic arrangements are dictated by the distances between nearest neighbours which are comparable to the order governing crystalline configurations. The structure factors $S(Q)$ corresponding to the X-ray scans of the RT and annealed films express the extent of scattering. These were derived using the raw intensity function and performing the corrections for background, polarization and absorption along with normalization. We used RAD program for this data processing (RAD, a program for analysis of X-ray diffraction data from amorphous materials [64]) and accordingly obtained the structure factors S(Q) for all the data sets. Fig. 3 compares the structure factors for all the films. Here too we see that the data corresponding to annealing temperature of 800 °C is fully crystalline, while others are partially or fully non-crystalline.

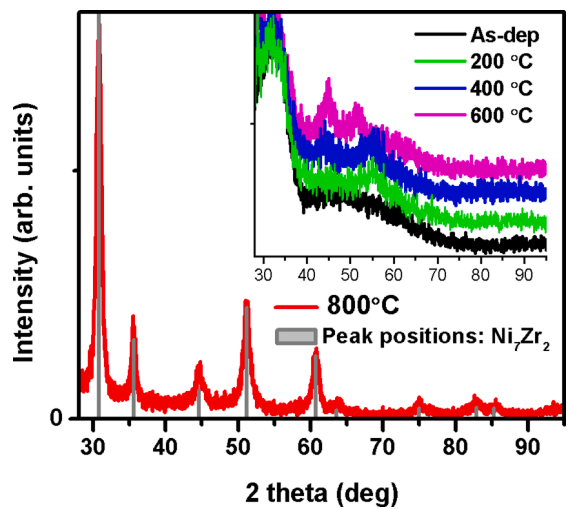

**Fig. 2.** GIXRD data of Ni-Zr alloy thin film annealed at 800 °C showing sharp crystalline peaks corresponding to the $Ni_7Zr_2$ phase. The inset shows amorphous nature of GIXRD plots recorded for as-deposited film and films annealed at 200 °C, 400 °C and 600 °C.

The pair distribution function (PDF) given by $g(r)$, is obtained from the Fourier transform of S(Q). It defines a probability density map of atomic distances $r$ between pairs of atoms in a material. The atoms are considered to be arranged on bonding spheres around a central atom. The derived PDF appropriately describes the probability of finding atoms at some distance $r$ from a central atom. It thus yields information about atomic structure and short range order of both amorphous and crystalline materials. The amorphous configurations are characterized by a few initial well-defined radii of bonding spheres, but further structures at higher $r$ are largely damped due to high overlaps between the distances. The function converges to unity signifying that the probability of finding an atom becomes 1 (one) for any distance. This is clearly seen in Fig. 4, which compares PDFs of the thin alloy films before and after annealing in the short range order regime. The crystalline film however exhibits sharper maxima which are sustained at higher $r$ as well, reflecting the symmetric arrangement of atoms.

The PDF peak positions are indicative of average interatomic distances and any shift in these peak positions can be interpreted as a measure of the change in bond length. Here the average distances between Ni and Zr atoms are represented by the crystallographic distance Ni-Zr. It was observed from the shifts in the first of the PDF peaks that the average bond lengths of the Ni-Zr alloy thin films differed by 0.14 Å with annealing. Variation in PDF peak widths of any system allows the determination of the extent of order/disorder introduced as a result of thermal annealing. It was found here that the PDF peak widths monotonically decreased with annealing temperature pointing to the control of structural disorder in the system and indicating that the system was being driven towards an ordered crystalline structure. Further, the increase in integral intensities of the PDF peaks due to rise in temperature pointed to the concomitant change in coordination number, and thus higher correlations between the atoms.

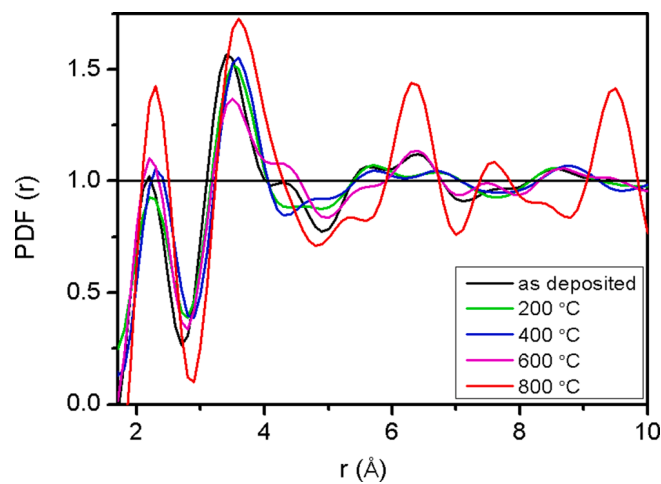

**Fig. 4.** Pair distribution functions calculated in the short range order regime for the as-deposited and annealed films.

The pair correlation function $G(r)$ estimates the distance between two particles in a random distribution, thus measuring how closely they are packed. In terms of PDF, it can be expressed as $G(r) = 4\pi\rho r(g(r)-1)$, where $\rho$ is the average atomic density in $Å^{-3}$ units. $G(r)$ is also related to S(Q) via a sine Fourier transformation. The radial distribution function (RDF) or $R(r)$, obtained from inverse Fourier transform of the diffraction data also measures similar aspects, but as the term suggests it gives the probability of the number of atoms in an annulus $dr$ at a distance $r$ from another atom. Its relation with PDF is given as $R(r) = 4\pi\rho r^2 g(r)$. Fig. 5 compares $R(r)$ for Ni-Zr alloy thin films before and after annealing. The RDFs for the films have been computed only up to 10 Å in order to focus attention on the changes occurring in the short range order features during devitrification of this system.

From XRR analyses of the films it is apparent that the contributions of 3 structures of $Ni_7Zr_2$, NiO and $ZrO_2$ relative to each other are to be considered for a complete description of the transformation due to annealing. The corresponding nearest neighbor distances of these structures can be associated with *R(r)* calculated for the films through a coordination analyses shown in Fig. 6. This figure shows histograms which are the *R(r)* functions for different neighbours in the $Ni_7Zr_2$ crystal structure.

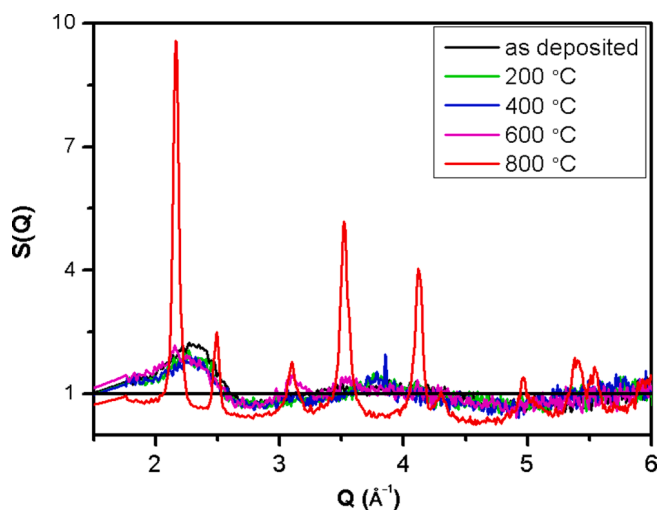

**Fig. 3.** Comparison of structure factor functions computed for as-deposited and annealed Ni-Zr alloy thin films.

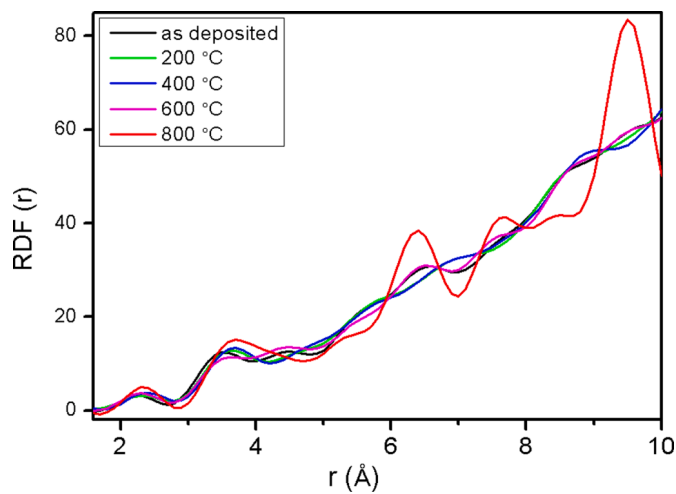

**Fig. 5.** Radial distribution functions calculated for as- deposited and annealed films. Due to experimental limitations only the first peak has been considered for coordination analyses.

Finally, the total distribution function *T(r)* used to describe the entire scattering results was calculated as $R(r)/r$ or $T(r) = 4\pi\rho r g(r)$. This is also related to the pair correlation function by $T(r) = 4\pi\rho r + G(r)$. This function is usually used to deconvolute the different distances in the data using symmetric peak functions.

It is pertinent to mention here that the spatial resolution of the RDF computed from the diffraction data is dependent on the maximum value of *q* used in the experiment. Hence in order to make these assessments in a completely error-free manner, it is preferred to have data recorded using much slower scans (to get high statistical accuracy of the intensities) and upto much higher *q* values, than those used in this work. Hence the analyses in this work have been restricted to the first observed peak and up to a radial distance of 3 Å only, which can yield reasonably good results. Within the range of r considered, the relevant atomic pairs of significance are as follows: Ni-O of NiO, Zr-O of $ZrO_2$ and Ni-Ni and Ni-Zr of $Ni_7Zr_2$. These are the nearest neighbours contributing to *R(r)*. The coordination analyses from Fig. 6 were used to identify various atomic pairs corresponding to the total correlation function for the alloy films before and after annealing. Fig. 7 shows this deconvolution of *T(r)* for each film state.

The total correlation function (*T(r)*) associated with each of the as-deposited and annealed alloy films are represented by cumulative fits (solid magenta lines in Fig. 7) to the respective data (open circles), which comprise contributions from the Ni-O, Zr-O, Ni-Ni and Ni-Zr atomic pairs. The deconvolution of *T(r)* corresponding to each temperature led to 3 components symbolized by red (dot), green (for as-deposited: short dash; for all annealed films: short dash dot) and blue (solid line) curves. The sequential positions of these curves at different bond distances determined the contribution of the constitutive atomic pairs. In case of the as-deposited film, the red (dot) and green (short dash) curves both encompassed contributions from Ni-O and Zr-O. But for all the annealed films only the red (dot) curves corresponded to Ni-O and/or Zr-O. The green (short dash dot) curves in these cases occurred at slightly higher *r* and hence represented the Ni-Ni bonds in $Ni_7Zr_2$. In case of the film annealed at 400 °C, Ni-Ni pairs were observed at lower *r* as well. Therefore only in this case, the red (dot) curve represents contributions from both Zr-O and Ni-Ni of $Ni_7Zr_2$. The blue (solid line) curve of the as-deposited film could be identified with contribution from the Ni-Ni pairs of the $Ni_7Zr_2$ structure. However, the blue (solid line) curves for all the annealed films corresponded to the position of the Ni-Zr atomic pair of the $Ni_7Zr_2$ structure. From these analyses, the composition of the Ni-Zr alloy films can be qualitatively described as $Ni_7Zr_2$ alloy with some contribution from oxide components.

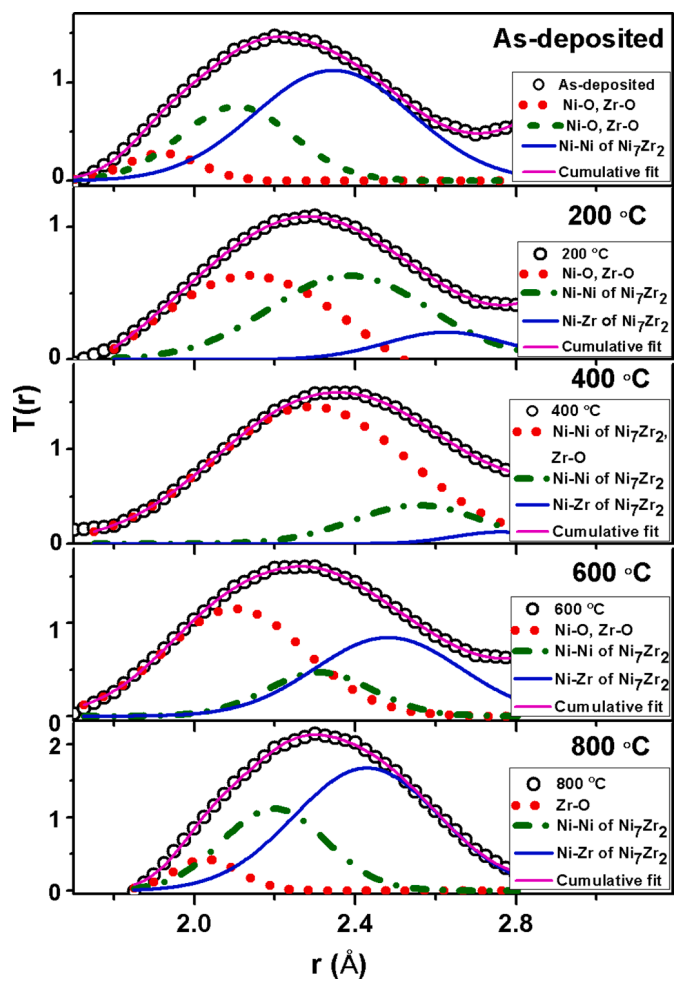

**Fig. 7.** Identification of various atomic pairs (Ni-O, Zr-O and both Ni-Ni and Ni-Zr from $Ni_7Zr_2$) contributing to the total correlation (T(r)) functions for as-deposited and annealed films.

### 3.3. Analysis of EXAFS data

EXAFS gives information regarding nearest neighbor distance,

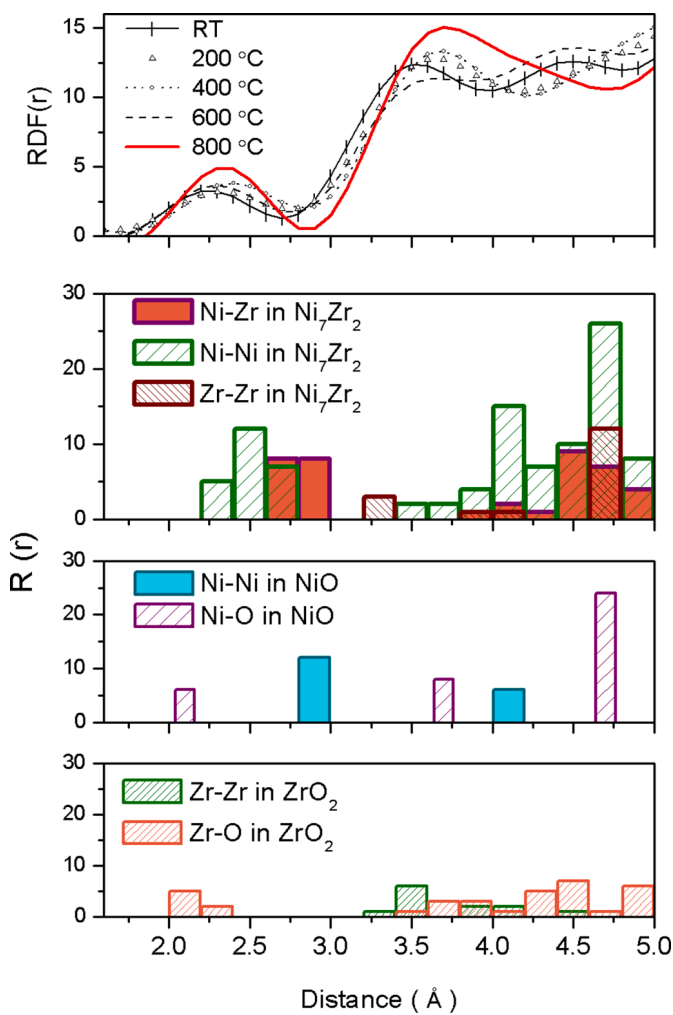

**Fig. 6.** R(r) corresponding to annealed films as well as crystalline $Ni_7Zr_2$, NiO and $ZrO_2$ respectively. Relevant atom-atom correlations Ni-O, Zr-O, Ni-Ni and Ni-Zr are shown separately for the crystalline samples.

coordination number and extent of disorder up to a maximum of three coordination shells surrounding a probe atom. Closely associated with this is another structural technique: X-ray absorption near edge structure (XANES). This is used to assess the oxidation states of the particular atomic species in the material and occupancy of the final bound states. EXAFS derives the partial RDF of the system as opposed to the total RDF derived through the procedure of total X-ray scattering analysis detailed in the previous section. EXAFS and XANES measurements were performed on the as-deposited and annealed Ni-Zr alloy thin films at K-absorption edges of both Ni and Zr in fluorescence mode. An advantage of using EXAFS for this study was that the absorption edges of both constituents fall into easily accessible spectral regions.

In an EXAFS measurement in fluorescence mode, the plot of absorption $E$ versus photon energy $\mu$ is obtained by scanning the required energy range and recording the incident intensity $I_0$ and the fluorescent intensity $I_f$ respectively and using the relation: $\mu = I_f / I_0$. The oscillations in the $\mu$ versus $E$ absorption spectrum are of real concern for EXAFS characterization. The data were therefore converted to $\chi$ versus $E$ data, with $\chi$ defined as [65]: $\chi(E) = (\mu(E) - \mu_0(E))/\Delta\mu_0(E_0)$, where $E_0$ is the absorption edge, $\mu_0(E)$ is the bare atom background, and $\Delta\mu_0(E_0)$ is the step in the $\mu(E)$ value at the absorption edge. The energy scale was subsequently converted to the wave number $k$ scale by: $k = ((2m(E - E_0))^{0.5})/h.\chi(k)$ is then weighted by $k^2$ to amplify the oscillations at high $k$ value and Fourier transformed to generate $\chi(r)$ versus $r$ in terms of real distances from the center of the absorbing atom. A set of EXAFS data analysis programs available within the IFEFFIT software package [66] have been used for reduction and fitting of the experimental EXAFS data.

The oxidation states of Ni and Zr in all the films were probed by XANES. Fig. 8(a) shows the normalised XANES spectra measured around Ni K-edge at 8333 eV for the as-deposited film and the films annealed at 400 °C, 600 °C and 800 °C. The XANES spectrum of a pure Ni metal foil is also shown in the figure for comparison. It is clearly seen that the Ni K-edge positions of the films annealed at 600 °C and 800 °C coincide with that of pure Ni foil, suggesting a metallic nature of Ni in these samples. However, the absorption edge of Ni in the as–deposited and the 400 °C annealed films were found to appear at higher energies suggesting the presence of NiO state in these films. The XANES spectra of these films also showed a sharp white-line like feature, characteristic of oxides. These results suggested the presence of NiO in the as-deposited Ni-Zr films, though the nature of the films turned metallic upon annealing at higher temperatures of 600 °C and 800 °C. The normalized XANES spectra measured around Zr K-absorption edge at 17,998 eV for the as-deposited Ni-Zr thin film and the films annealed at 400 °C, 600 °C and 800 °C are compared with that of pure Zr metal foil in Fig. 8(b). The incidence of $ZrO_2$ was detected in the films, but its presence diminished as the films were annealed to higher temperatures.

*3.3.1. EXAFS analyses at Ni absorption edge*

The radial distribution $\chi(r)$ versus $r$ plots for the as-deposited and annealed Ni-Zr alloy films, which were derived from the experimental $\mu$ versus $E$ plots as explained above, are shown in Fig. 9. The fourier transform of the raw data was carried out in the $k$ range of 2–9 $Å^{-1}$. The solid curves in the figure are the best fit theoretical curves. The fitting range in $r$ space was decided in such a way that the number of free variables used during fitting was always below the upper limit of $(2\Delta k\Delta r/\pi + 1)$, set by the Nyquist criterion [67]. The data range fitted up to 3 Å in this manner is shown in Fig. 9, and the data beyond this value which could not be fitted is shown as an inset in this figure. The Ni-Ni and Ni-Zr pathways of $Ni_7Zr_2$ are compared with the data recorded for the films at higher $r$ in the inset.

It can be seen from Fig. 9 that the $\chi(r)$ versus $r$ plots corresponding to the as-deposited film and the film annealed at 400 °C are similar to each other but are completely different from that of the films annealed at 600 °C and 800 °C. Taking into consideration the evidence obtained from the relevant XANES analyses, the data of both the as-deposited film and the film annealed at 400 °C were fitted with NiO structure. The structural parameters of NiO were taken from the Inorganic Crystal Structure Database (ICSD) website [68]. The theoretical simulation of the spectrum for the film annealed at 600 °C was found to fit the data best on consideration of a mixed state comprising NiO and $Ni_7Zr_2$. The parameters of $Ni_7Zr_2$ structure used for the simulation were taken from reference 49. Finally, the radial distribution plot for the film annealed at 800 °C fitted well with the crystalline phase $Ni_7Zr_2$, which matched with the phase identification through GIXRD measurement on this film (see

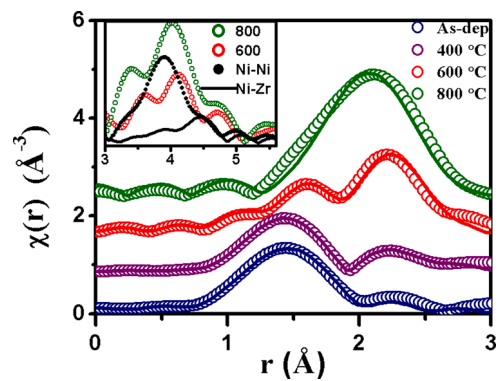

**Fig. 9.** EXAFS measurements made at Ni absorption edge of as-deposited and annealed films. Open circles represent data points and solid lines are model fits to the data. Inset shows scans made at higher radial distances for 600 °C and 800 °C annealed films and compared with theoretical plots of $Ni_7Zr_2$ path structures.

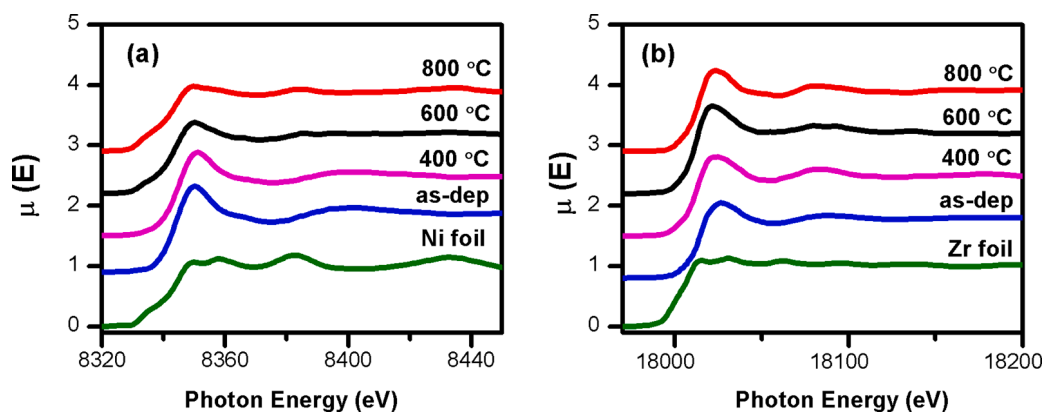

**Fig. 8.** Normalized XANES spectra of the as-deposited and annealed Ni-Zr alloy thin films measured at (a) Ni absorption K-edge and (b) Zr absorption K-edge. The traces have been vertically displaced for clarity.

Fig. 3). The best fit parameters obtained from fitting of EXAFS data for all the films are given in Table 2.

The oxygen coordination around a Ni atom was seen to decrease from 4.97 to 2.33 when the as-deposited film was annealed up to 600 °C, which showed that the oxide contribution in the films reduced with rising temperature. There was no existence of NiO in the film annealed at 800 °C. In the theoretically generated $Ni_7Zr_2$ model, the bond length of the first nearest neighbor Ni-Ni was 2.33 Å with coordination number (CN) of 8 and that of the second nearest neighbor Ni-Zr was 2.79 Å with CN of 4. These coordinations were observed in the TFMGs annealed at higher temperatures. As the films were annealed from 600 °C to 800 °C both Ni-Ni and Ni-Zr bond lengths were seen to decrease. The corresponding coordination numbers also increased with annealing temperature: for the Ni-Ni pair, the coordination numbers increased from 3.28 to 4.98 and for the second nearest neighbor, Ni-Zr, the coordination numbers increased from 2.16 to 2.49. The increase in coordination is representative of an increase in the extent of short range order in the system as a result of annealing. The films annealed at 600 °C and 800 °C thus displayed increasing coordination of both Ni-Ni and Ni-Zr pairs accompanied by shortening of corresponding bond lengths, indicating the growth of the $Ni_7Zr_2$ phase with annealing.

*3.3.2. EXAFS analyses at Zr absorption edge*

The FT-EXAFS spectra recorded at Zr absorption edges of the as-deposited and annealed Ni-Zr films along with respective theoretical fits to the data are shown in Fig. 10. The data of the films at R.T. and 400 °C could be fitted with a theoretically generated $ZrO_2$ structure.

The major peak occurring in the data was found to correspond to the nearest oxygen shell (Zr-O) from the central Zr atom of the $ZrO_2$ structure. The simulation of the FT-EXAFS spectra was generated assuming the oxygen shell (Zr-O) at 2.04 Å with a CN of 5 corresponding to bulk $ZrO_2$ structure. The parameters of $ZrO_2$ structure were taken from the Inorganic Crystal Structure Database (ICSD) website [68]. The best fits applied to the data of the films annealed at 600 °C and 800 °C were obtained with a mixed phase ($ZrO_2$+ $Ni_7Zr_2$) model. The presence of $Ni_7Zr_2$ phase in these films is manifested by the appearance of peaks near 3 Å for these samples as shown in Fig. 10. The values of the parameters obtained for the best fits are listed in Table 3.

The oxygen coordination in the thin alloy films appeared to decrease as temperature was increased, with an anomalous rise at 400 °C. This confirmed the depletion of the oxide phase with annealing. The associated Zr coordination in the second shell (Zr-Zr) were seen only in the films annealed at 600 °C and 800 °C and were also found to be slightly more than the theoretical value. This may be attributed to contributions arising from higher Zr-Zr shells of the $Ni_7Zr_2$ phase; revealing lower contributions of $ZrO_2$ phase in these two films. The coordination of the Zr-Ni pairs associated with the $Ni_7Zr_2$ structure, from both the first and second nearest neighbours, also rose when the films were annealed from 600 °C to 800 °C. This further signified higher contribution of the $Ni_7Zr_2$ phase compared to the $ZrO_2$ phase, as expected from the results of analyses by other techniques on these films explained in previous sections.

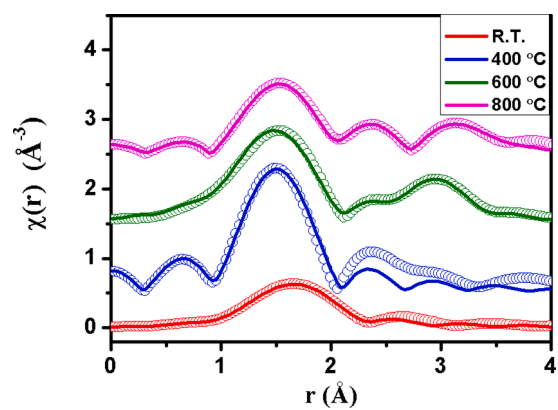

**Fig. 10.** EXAFS measurements made at Zr absorption edge for as-deposited and annealed films. Open circles represent data points and solid lines are model fits to the data.

## 4. Discussion

The investigation of the thermal evolution of annealed Ni-Zr alloy thin films through the techniques detailed in this work indicated the presence of an underlying $Ni_7Zr_2$ structure accompanied by a lesser degree of associated oxides. The relative amount of oxide phases decreased with increasing annealing temperature as seen from the consistent analyses of XRR, GIXRD and EXAFS data. The transition from amorphous to final crystalline state took place through a polymorphous transformation undeterred by the presence of oxygen in the initial stages, unlike the change in crystallization pathway due to oxygen reported for other Ni-rich Ni-Zr alloys [69]. Densification of the films also increased with annealing temperature as seen from the ESLD profiles. The slight increase of Zr oxide contribution at 400 °C observed in the XRR model analyses was also confirmed through EXAFS measurements, which is why there was no significant volume change seen between the ESLD plots of as-deposited and 400 °C films. Deconvolution of the total correlation function obtained from GIXRD whole pattern analyses at each annealing temperature also indicated the existence of a base $Ni_7Zr_2$ structure, through the presence of Ni-Ni and Ni-Zr bonds. The decrease of peak widths with annealing indicated a decrease in the structural disorder of the system, as it transitioned from amorphous to crystalline state. The consequent change in average coordination properties was

**Table 2**

EXAFS fitting results of Ni-Zr thin alloy films at Ni absorption edge. Structural parameters include: *r* (interatomic distance), *CN* (coordination number), *σ* (relative displacement of atoms). The goodness of fit is defined by the value of R-factor.

| Fitted paths | | As-deposited | 400 °C | 600 °C | 800 °C |
|---|---|---|---|---|---|
| NiO structure | | | | | |
| Ni-O | CN(6) | 4.97 ± 0.36 | 2.88 ± 0.09 | 2.33 ± 0.27 | – |
| | r (2.08) | 1.99 ± 0.03 | 1.96±0.01 | 2.02 ± 0.01 | – |
| | $\sigma^2$ | 0.008 ± 0.001 | 0.002 ± 0.001 | 0.014 ± 0.002 | – |
| Ni-Ni | CN(12) | – | 1.02 ± 0.11 | – | – |
| | r (2.94) | – | 2.85 ± 0.01 | – | – |
| | $\sigma^2$ | – | 0.002 ± 0.001 | – | – |
| $Ni_7Zr_2$ structure | | | | | |
| Ni-Ni | CN(8) | – | – | 3.28 ± 0.37 | 4.98 ± 0.56 |
| | r (2.33) | – | – | 2.45±0.01 | 2.39 ± 0.01 |
| | $\sigma^2$ | – | – | 0.002 ± 0.001 | 0.004 ± 0.001 |
| Ni-Zr | CN(4) | – | – | 2.16 ± 0.06 | 2.49 ± 0.28 |
| | r (2.79) | – | – | 2.76 ± 0.01 | 2.74 ± 0.01 |
| | $\sigma^2$ | – | – | 0.002 ± 0.001 | 0.007 ± 0.001 |
| R-Factor | | 0.003 | 0.0011 | 0.007 | 0.004 |

**Table 3**
EXAFS fitting results of Ni-Zr thin alloy films at Zr absorption edge. Structural parameters include: r (interatomic distance), CN (coordination number), σ (relative displacement of atoms).

| Fitted paths | | As-deposited | 400 °C | 600 °C | 800 °C |
|---|---|---|---|---|---|
| $ZrO_2$ structure | | | | | |
| Zr-O | CN(5) | 4.62± 0.09 | 5.72±0.32 | 4.2 ± 0.47 | 3.74±0.46 |
| | r (2.04) | 2.08± 0.01 | 2.09±0.01 | 2.12±0.01 | 2.12±0.01 |
| | $\sigma^2$ | 0.015 ±0.004 | 0.003 ±0.001 | 0.012 ± 0.001 | 0.002 ±0.001 |
| Zr-Zr | CN(3) | – | – | 3.75 ± 0.52 | 3.27±0.21 |
| | r (3.35) | – | – | 3.36±0.01 | 3.48± 0.01 |
| | $\sigma^2$ | – | – | 0.002 ± 0.001 | 0.01± 0.001 |
| $Ni_7Zr_2$ structure | | | | | |
| Zr-Ni | CN(8) | – | – | 4.64 ± 1.36 | 6.0 ± 0.12 |
| | r (2.75) | – | – | 2.93 ± 0.08 | 3.02±0.02 |
| | $\sigma^2$ | – | – | 0.023 ±0.018 | 0.02± 0.003 |
| Zr-Ni | CN(6) | – | – | 4.56± 1.02 | 6.18±1.68 |
| | r (2.91) | – | – | 3.08±0.06 | 3.12±0.03 |
| | $\sigma^2$ | – | – | 0.015 ±0.003 | 0.02±0.01 |
| R-Factor | | 0.0002 | 0.002 | 0.011 | 0.0027 |

also seen to be reflected in the variation of integral intensities corresponding to the PDF peaks. The average bond length variation obtained from PDF analyses over all annealing temperatures was ~0.14. This corroborated with the change in $Ni_7Zr_2$ bond lengths obtained from EXAFS measurements, which was well within this range. Analyses of the coordination environment of each atom derived from analyses of EXAFS spectra provided information regarding the chemical short range order in this metallic glass system undergoing devitrification [70]. At the Zr absorption edge, comparison of the Ni coordination around Zr atoms when the films were annealed from 600 °C to 800 °C, indicated a 29% rise in the first shell of nearest neighbours at radial distance of 2.75 Å and 35% increase in the second shell at 2.91 Å. In contrast, from the EXAFS analyses at the Ni edge, the Zr coordination around Ni increased from 600 °C to 800 °C by a comparatively lesser extent of 15%. This trend can be attributed to the atomic size difference between Ni and Zr which prompts better arrangement of Ni atoms around Zr atoms than the converse, as reported in Ni-rich Ni-Zr alloy thin films [30]. The corresponding Ni-Ni and Ni-Zr bond lengths were seen to decrease along with increasing coordination numbers as the devitrification progressed with annealing. Such structural changes accompanying chemical ordering have also been reported in bulk Ni-Zr alloys during the transformation of amorphous to crystalline states, wherein the role of short range order is evident [56].

In case of oxide contributions seen at both absorption edges, the coordination environment of both Ni-oxide and Zr-oxide appeared to get sparser with increasing annealing temperature. Respective bond lengths were also seen to increase with devitrification. At Ni absorption edge the coordination of Ni-O that was obtained for the as-deposited film was seen to drop by 42% when the film was annealed at 400 °C and by 53% at 600 °C. At the final annealing temperature of 800 °C wherein $Ni_7Zr_2$ was the dominant phase, the Ni oxide contribution was completely absent. On the other hand, at the Zr absorption edge, the contribution of Zr-oxide initially increased slightly at 400 °C by 23.8% followed by a steady decrease thereafter by 26.6% up to 600 °C and by 34.6% at 800 °C. The attainment of the final crystallized state thus proceeded notwithstanding the presence of oxides, much the same as in the case of Ni rich bulk amorphous Ni-Zr alloy [71].

## 5. Summary and conclusions

The local atomic information derived at various annealing stages during devitrification of magnetron sputter deposited Ni-Zr alloy thin films pointed to the predominant formation of the $Ni_7Zr_2$ structure. Detailed analyses of the films at each stage through XRR, GIXRD and EXAFS techniques effected evolution of local atomic structure during the devitrification process. modeling of XRR data showed that film thickness decreased with annealing while densification increased. The traces of NiO and $ZrO_2$ present in the initial stages, did not act as deterrents to the polymorphous transformation during annealing of the Ni-Zr alloy films. The oxide content detected in the films was also noted to decrease rapidly as annealing proceeded; leaving the crystallization pathway unhindered. As expected for the thermal stability of such metallic glass systems, the chemical short range order played a major role. Applying a whole pattern analysis to the GIXRD data revealed decrease in pair distribution function peak widths with annealing, indicating a control over structural disorder in the system as it transitioned from the amorphous (disordered) to crystalline (ordered) state. The bonding of unlike pairs of atoms (Ni-Zr) took precedence over that of like atoms (Ni-Ni, Zr-Zr) while the films developed structural order. The consequent change in average coordination properties was also seen to be reflected in the variation of integral intensities corresponding to the radial distribution function. Partial atomic distribution in the environment of each constituent was examined at every stage of annealing through EXAFS measurements, which gave proper insight into atomic scale changes to the order present.

Thus we have been able to report the evolution of short range order in the Ni-Zr TFMG as a function of annealing temperature. The investigative techniques used in this study all provided coherent and corroborative results which gave a detailed overall view of the devitrification process.

**Credit author statement**

Debarati Bhattacharya: Conceptualization, Investigation, Methodology, Formal analysis, Validation, Visualization, Writing - Original Draft, Writing - Review & Editing.
Nidhi Tiwari: Methodology, Formal analysis, Validation.
P.S.R. Krishna: Methodology, Formal analysis, Validation.
Dibyendu Bhattacharyya: Validation, Review.

**Declaration of Competing Interest**

The authors declare that they have no known competing financial interests or personal relationships that could have appeared to influence the work reported in this paper.

**Data Availability**

Data will be made available on request.

**Acknowledgments**

The author thanks Swapan Jana and Pooja Moundekar for their technical help during thin film deposition and X-ray measurements and Dr. S.L. Chaplot for informative discussions. The author is grateful to (late) Dr. C.L. Prajapat for annealing of the thin films.